# The Second Law of Thermodynamics, Life and Earth's Planetary Machinery Revisited

Axel Kleidon
Max-Planck-Institute for Biogeochemistry
Hans-Knöll-Str. 10
07745 Jena, Germany

e-mail: akleidon@bgc-jena.mpg.de
web: http://gaia.mpg.de

**Abstract**

Life is a planetary feature that depends on its environment, but it has also strongly shaped the physical conditions on Earth, having created conditions highly suitable for a productive biosphere. Clearly, the second law of thermodynamics must apply to these dynamics as well, but how? What insights can we gain by placing life and its effects on planetary functioning in the context of the second law? In Kleidon (2010), I described a thermodynamic Earth system perspective by placing the functioning of the Earth system in terms of the second law. The Earth system is represented by a planetary hierarchy of energy transformations that are driven predominantly by incoming solar radiation, these transformations are constrained by the second law, but they are also modified by the feedbacks from various dissipative activities. It was then hypothesised that life evolves its dissipative activity to the limits imposed by this hierarchy and evolves feedbacks aimed at pushing these limits to higher levels of dissipative activity. Here I provide an update of this perspective. I first review applications to climate and global climate change to demonstrate its success in predicting magnitudes of physical processes, particularly regarding temperatures, heat redistribution and hydrological cycling. I then focus on the limits to dissipative activity of the biosphere. It would seem that the limitations by thermodynamics act indirectly by imposing limitations associated with transport and material exchange. I substantiate this interpretation and discuss the broader implications for habitability, the emergence and evolution of life, and the contemporary biosphere.

## 1. Thermodynamics, Life, and Earth: More than the sum of its parts

A thermodynamic interpretation of life has a long history in science. In the context of the Earth system, the first notable interpretation is probably that of Boltzmann (1886), who wrote that life uses the large temperature difference between the hot Sun and the cold Earth. Since then, several authors have extended such a thermodynamic interpretation of life. Lotka (1922a, b) expressed life's evolution by natural selection as an evolution towards a state of maximum power, while Schroedinger (1944) clarified that life follows the second law by maintaining its metabolism from the entropy difference between low entropy food and high entropy waste. Odum and Pinkerton (1955) applied this line of interpretation to ecosystems and linked this to possible applications of a maximum power limit, while Lovelock and Margulis (1974) extended this line of reasoning to the planetary scale in form of the proposed Gaia hypothesis. This line of research continues in ecological research to this day, as reflected in the recent reviews by Nielsen et al. (2020) and Hall and McWhirter (2023).

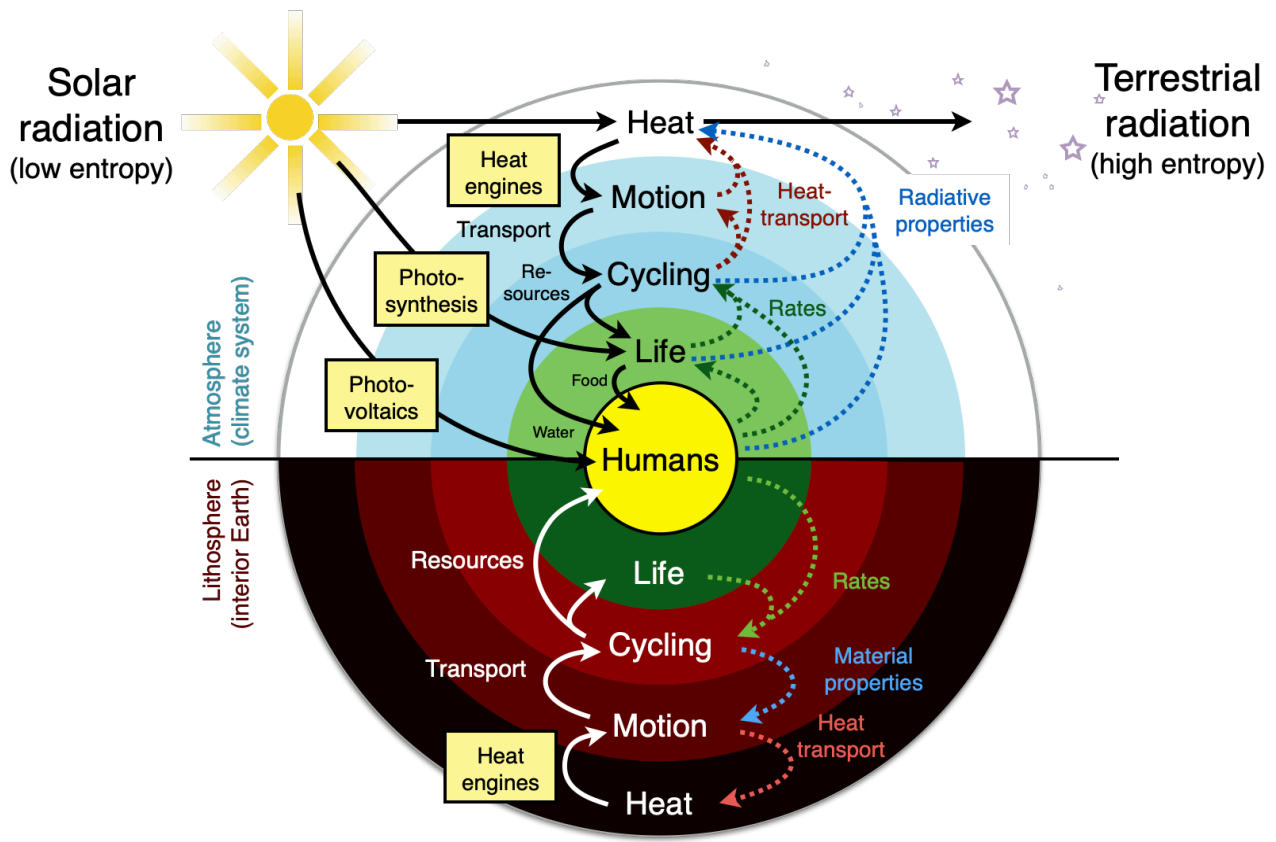

*Figure 1:* *The Earth system as a hierarchy of energy transformations, with yellow boxes illustrating the processes that perform work from the primary forcings. The physical environment in form of motion and cycling set constraints to life, yet life also contributes to these dynamics by generating chemical free energy by photosynthesis (yellow box) and by altering rates and radiative properties of the environment (dashed arrows). Taken from Kleidon (2023a), after Kleidon (2010) and Kleidon (2016).*

So it is clear that the second law of thermodynamics applies to life. This may seem like a trivial insight. After all, all Earth system processes must follow the second law of thermodynamics. This leads us to the question whether there is more to learn about life from the second law. In Kleidon (2010), I suggested a clear "yes" to this question, but argued that the answer is a bit more complex because the second law is manifested in more indirect ways in overall Earth system functioning: Entropy constrains how much work can be derived from heat, as reflected in the well-established Carnot limit of a heat engine, work is needed to sustain motion, motion and transport is needed for material cycling, and life depends and interacts with this material cycling as it needs to maintain the exchange of reactants and products associated with its various metabolisms.  In other words, it is less about entropy itself and how it is increased, but more about how it limits which kinds of work are done in the Earth system, the consequences of this work, and how life affects and interacts with these processes that perform work.

To start, I began by describing the Earth system as a sequence of energy transformations, starting with solar radiation as the source of energy with very low entropy, its subsequent transformations into heat, motion, the hydrological cycle, and other Earth system processes.  This is shown schematically in Figure 1 (see also Kleidon, 2016, 2023a).  Each transformation follows the second law of thermodynamics.  It sets a clear causal direction as well as limits on how much work can be derived from sunlight, and thus on how dissipative the Earth system can be.  Understanding these limits requires thermodynamics, but also the context of the Earth system, because interactions with boundary conditions (as indicated by the dashed arrows in Figure 1) play a critical role in setting these limits.  Life depends on these processes because they are intimately connected to the availability of resources, but life also changes these processes by its activity, altering material cycles, chemical compositions, and radiative properties (dashed arrows in Figure 1), which in turn affects the conditions in which life maintains its activities.  To understand life as a planetary phenomenon, it thus requires more than the sum of thermodynamics, life, and its environment because of the critical importance of interactions.

This contribution aims to provide an extension of this perspective, making the application of thermodynamics and related constraints to life on Earth more specific.  To do so, I first describe this hierarchy of energy transformations in more detail and illustrate the important role of interactions in setting relevant limits.  I do this to demonstrate that this approach has been highly successful in deriving first-order estimates for climate and global climate change.  This substantiates that the connections drawn between the outer shells in Figure 1 represent the first-order causes and effects that shape these Earth system processes.  To develop an equally successful approach to understand life, one then needs to first identify the relevant constraints of life in the context of this thermodynamic Earth system perspective.  This is what I aim to do in Section 3, demonstrating that it appears that thermodynamic limits relate primarily to constraints imposed on maintaining mass transport and exchange, rather than on energy conversions directly. I then illustrate that life appears to evolve to minimise frictional losses and pushes the imposed transport constraints to higher levels by linking this view to some established concepts and previous literature.  I close with an outline of potential implications for understanding life as a planetary phenomenon.

**2. Thermodynamics and Earth system interactions shape transport limitations**

I first describe in greater detail by how much the combination of thermodynamics and interactions limit Earth system processes, making them predictable.  This concerns the outer layers of the "*Earth onion*" shown in Figure 1, from the radiative fluxes to heating to motion and cycling, with work being performed by heat engines.  The Earth is unevenly heated by solar radiation, with the surface and the tropics absorbing more solar radiation than the atmosphere and the poles.  This results in heating differences, and these serve as the drivers to generate motion, just as it is the case for a heat engine.  For large-scale motion in the atmosphere, the warmer tropics serve as the heat source of the engine, the colder poles act as the cold sink, while the work generated by the engine is used to maintain motion against surface friction.  The same picture can be applied to the vertical, where the warmer surface acts as the heat source of a convective engine, while the cooler atmosphere represents the cold sink.

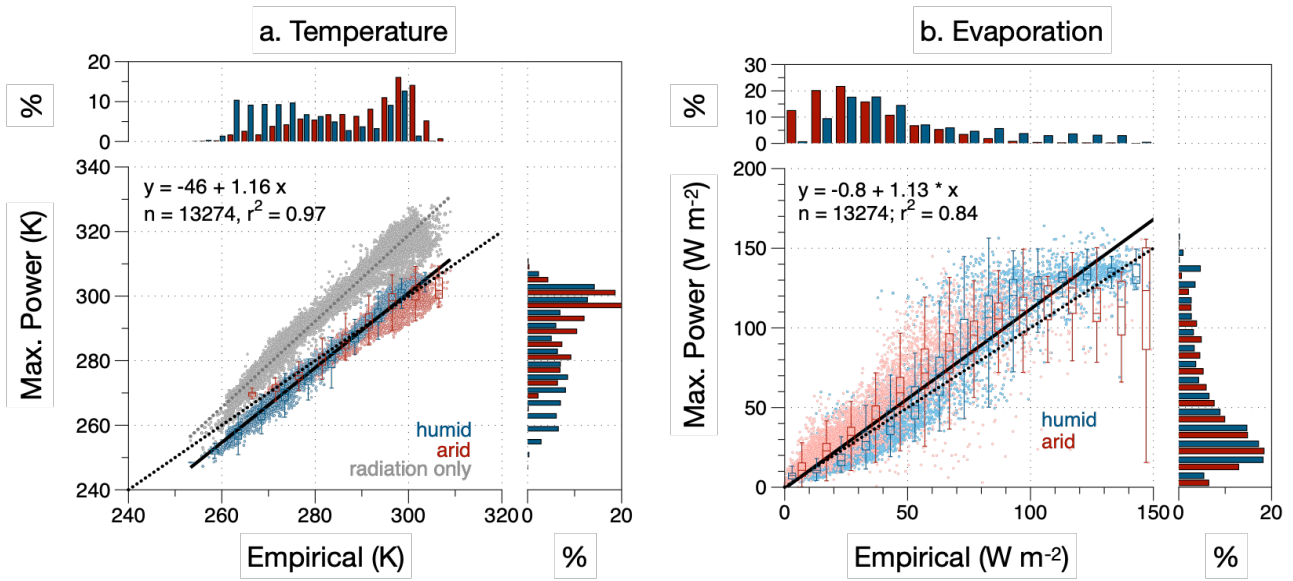

*Figure 2: Climatological (a.) mean temperatures and (b.) evaporation rates (in terms of the latent heat flux, i.e., multiplied by the latent heat of vaporisation) estimated from the maximum power approach compared to observations. The different dots represent each land grid point of global datasets, the grey dots in (a.) represent the temperatures in the absence of convective motion. From Kleidon (2021b).*

When this heat engine performs work and generates motion, it inevitably transports heat. This reduces the heating differences from which the work is generated from, and thereby levels out temperature differences, altering the boundary conditions of the engine. This depletion results in a maximum power limit to motion at which the heating difference is depleted to about half the value that would be attained in the absence of motion (Kleidon and Renner, 2013a; Kleidon 2021a). This power is eventually dissipated by friction. By limiting the power, the second law then sets the limit to the dissipative activity of the atmosphere and constrains the magnitude of material transport and exchange.

The limit can be applied to global datasets of the radiative forcing to yield estimates of heat fluxes and temperatures that compare very well with observations on land (Kleidon et al. 2014, Kleidon and Renner 2018, Ghausi et al. 2023). An example is given in Figure 2, showing climatological mean estimates for temperature and evaporation obtained from the maximum power limit being compared to global datasets (Kleidon 2021a). This approach also yields sensitivities to global climate change that are consistent with global climate models (Kleidon and Renner 2013b, Kleidon and Renner 2017), and the power associated with the limit constrains the availability of wind energy at the planetary scale (Miller et al., 2011; Miller and Kleidon 2016; Kleidon 2021b).

What this means is that the second law - in form of the Carnot limit of a heat engine - in combination with the uneven heating by the Sun as well as the interactions - in form of the resulting heat transport - set a predictive limit to motion in the atmosphere. The consistency with observations implies that the atmosphere works as hard as it can to generate motion: it works at its limit. It is the combination of thermodynamics and interactions that set this limit. This emphasises two aspects: that interactions are a critical component shaping this limit in the Earth system, and that the dissipative activity of the vastly complex atmosphere becomes predictable due to the second law.

This limit on motion then results in further, indirect constraints related to transport. Vertical motion maintains the mass exchange between the surface and the atmosphere, setting a limit to how much water can evaporate and simultaneously be brought into the atmosphere. In fact, the combination of the maximum power limit applied to vertical heating differences with the thermodynamic equilibrium settings of saturated air at the surface set an upper limit for evaporation, and thus to the intensity of the hydrological cycle. The evaporation rate that is derived from this combination compares very well to observations when soil water availability does not restrict evaporation rates (Kleidon and Renner 2013a, Kleidon et al. 2014, Conte et al. 2019**,** Kleidon 2021a). This concerns the link between motion and cycling shown in Figure 1. The estimates derived in this way do not represent a direct application of the second law nor of a maximum power limit, but rather an indirect manifestation by constraining transport. Thermodynamics thus restricts motion, and motion imposes limitations to other Earth system processes (Kleidon 2023a), and this applies to life as well.

**3. Life is indirectly limited by thermodynamics through transport**

To understand how this thermodynamic Earth system perspective applies to life, we first need to describe the biosphere as a thermodynamic Earth system and identify how this is linked to motion and transport as a limitation. The biosphere represents an open thermodynamic system composed of producers and consumers that exchange energy and mass with their environment (Figure 3). The term "producers" refer to those organisms that produce carbohydrates (notably, photosynthesisers such as plants and algae) versus those that consume it (heterotrophs, such as animals, although producers also need some carbohydrates to maintain their metabolism - as indicated by the dotted arrows in Figure 3). Thermodynamically speaking, the producers generate chemical free energy that is associated with the chemical disequilibrium of reduced, organic carbon compounds such as carbohydrates or hydrocarbons, and oxygen, while the consumers burn calories, that is, they deplete this disequilibrium with their metabolisms, turning the chemical free energy into heat and respiring carbohydrates and oxygen back into carbon dioxide and water.

What, then, limits the dissipative activity of the biosphere? The dissipative activity of the biosphere is ultimately constrained by how much chemical free energy is generated by photosynthesis and how much of this results in the formation of new biomass that can serve as food for the consumers. The question regarding the relevant limits is thus a question of what limits photosynthesis. This question has long been evaluated using thermodynamics (starting by Duysens, 1962, and, e.g., reviewed by Landsberg and Tonge, 1980), with the outcome being that the energy transformation from light to carbohydrate appears to only operate near its thermodynamic limit at low light concentrations (Hill and Rich, 1983). Natural ecosystems, however, typically operate with less than 3% efficiency (Figure 4), being much less than a thermodynamic efficiency of 18% or so inferred from energy conversions. So it would appear that the biosphere does not operate at its thermodynamic limit.

Yet, when one looks at photosynthetic carbon uptake differently, one can identify a relevant limit at which terrestrial ecosystems appear to operate (Kleidon 2021a) and that links to the Earth system view depicted in Figure 1. Besides light, photosynthesis needs carbon dioxide to store the energy in longer-lived organic carbon compounds. Carbon dioxide enters plants by the gas exchange through their stomata, through which they take up carbon dioxide from air while releasing water vapour. This gas exchange takes place at a relatively fixed ratio, resulting in a close correlation between photosynthetic efficiency and evaporation (Figure 4b). Evaporation is, in turn, limited by the vertical exchange with the atmosphere if water is sufficiently available, which in turn can be predicted from the maximum power limit on maintaining motion, as shown in Figure 2.

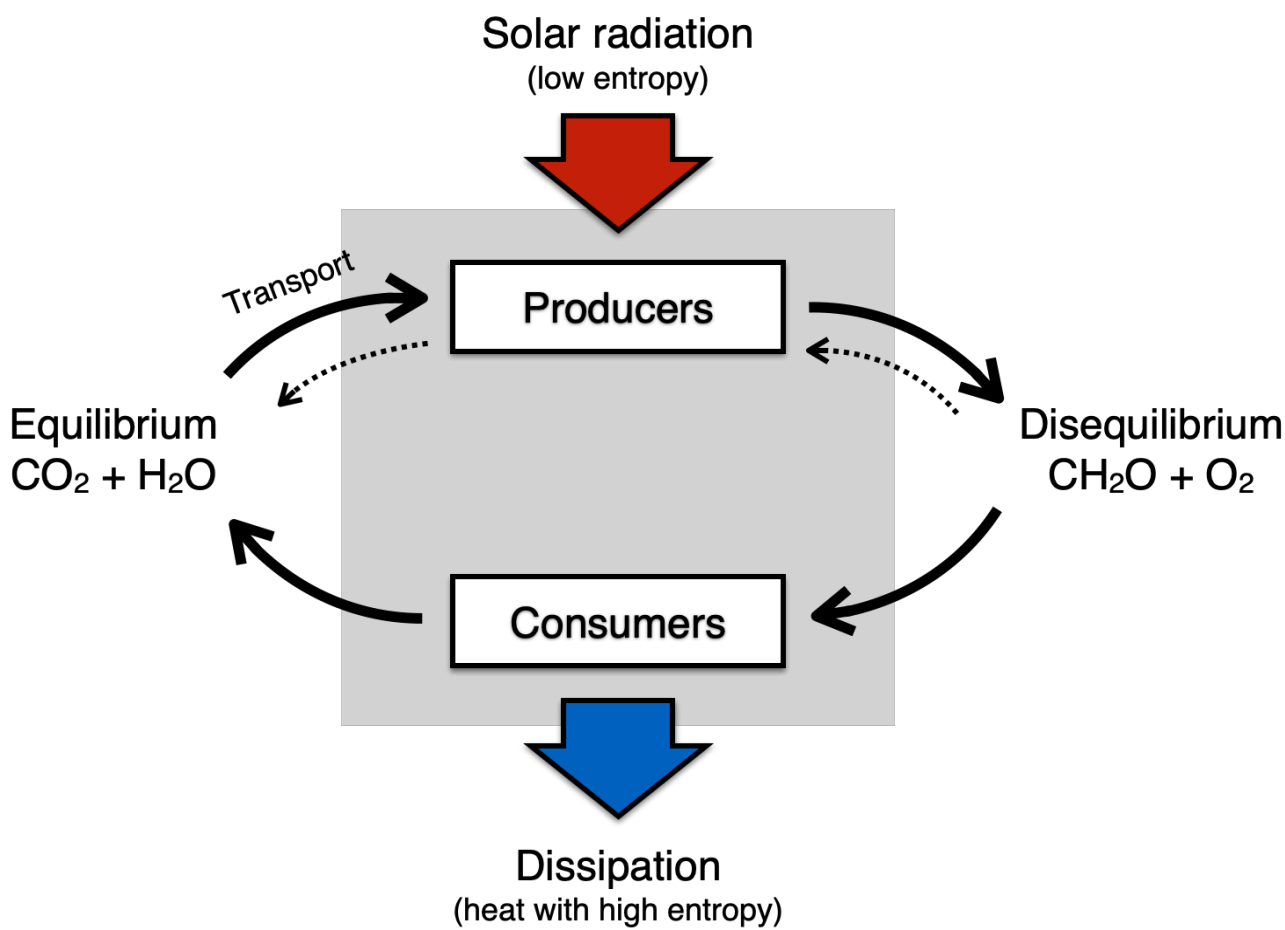

*Figure 3:* Schematic diagram of the biosphere as an open dissipative system composed of producers and consumers that is associated with generating chemical disequilibrium in form of reduced carbon compounds ($CH_2O$) and oxygen ($O_2$) and depleting it by dissipative activities through metabolisms into carbon dioxide ($CO_2$) and water ($H_2O$).

This yields an explanation of how thermodynamics indirectly limits the photosynthetic activity of the biosphere. This limitation does not concern the direct conversion of light into chemical free energy, but acts indirectly, as it limits transport and thus the rate of supplying reactants and removing products during the photosynthetic process. This, in turn, sets a physical limit to how much power the producers can provide to the dissipative activity of the biosphere.

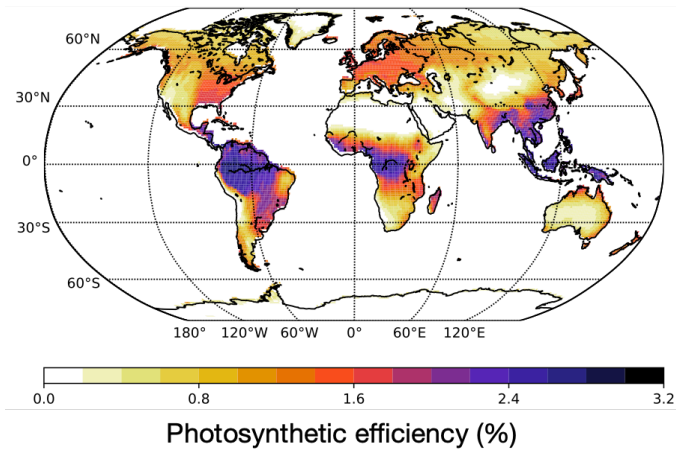
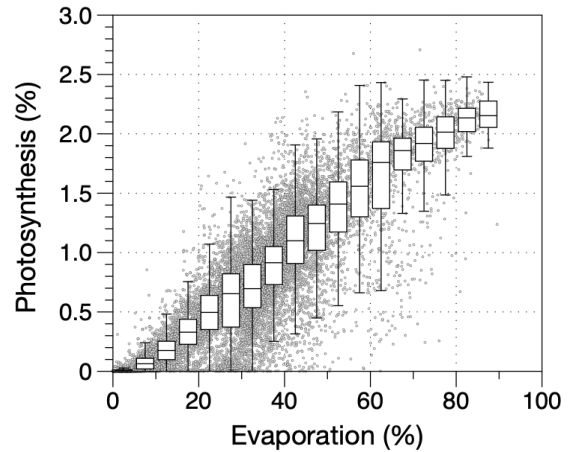

*Figure 4:* *(a.) Estimated climatological mean photosynthetic efficiency from satellite observations on land and (b.) its correlation with evaporation, expressed in terms of the energy equivalent of evaporation, the latent heat flux, as percentage of absorbed solar radiation at the surface. After Kleidon (2021a).*

**4. Life pushes limits by reducing resource and transport limitations**

The limitations that thermodynamics imposes on life at broader scales thus seem to be indirect, as they relate to transport limitations that the environment imposes upon them, and not on the energy conversions directly. This does, however, not imply that the biosphere is passively constrained by these limits. The biosphere has various means to improve transport mass within organisms, within communities, and to affect the environment in ways to push these limits to higher levels, allowing for greater power and dissipative activity. These ultimately need to be understood in the context of broader Earth system functioning (Figure 5). In the following, I want to provide a few examples to illustrate such means to optimise transport and enhance power of the biosphere.

To start, plants have optimised their way of transporting mass within themselves as well as how they assemble in communities. West et al. (1997) showed that the fractal, tree-like networks of plants result from the minimisation of frictional dissipation within their vascular networks that connect a spatially distributed resource within a volume to one point by a network of conduits. This fractal scaling of vascular networks then leads to the well-known scaling relationships of metabolic rate with organism size, known as Kleiber's law (Kleiber, 1932). This law states that metabolism - and thus dissipation as well as the exchange of reactants and products - scales with the body mass to the power of 3/4. Note that this scaling only applies to multicellular organisms and not to individual bacteria, where the metabolic rate scales linearly (Hoehler et al. 2023). Minimising frictional dissipation within their vascular networks implies that plants achieve the most transport for a given driving difference in the environmental forcing. And since transport limits photosynthesis, this optimised design would translate into a maximisation of photosynthetic capacity.

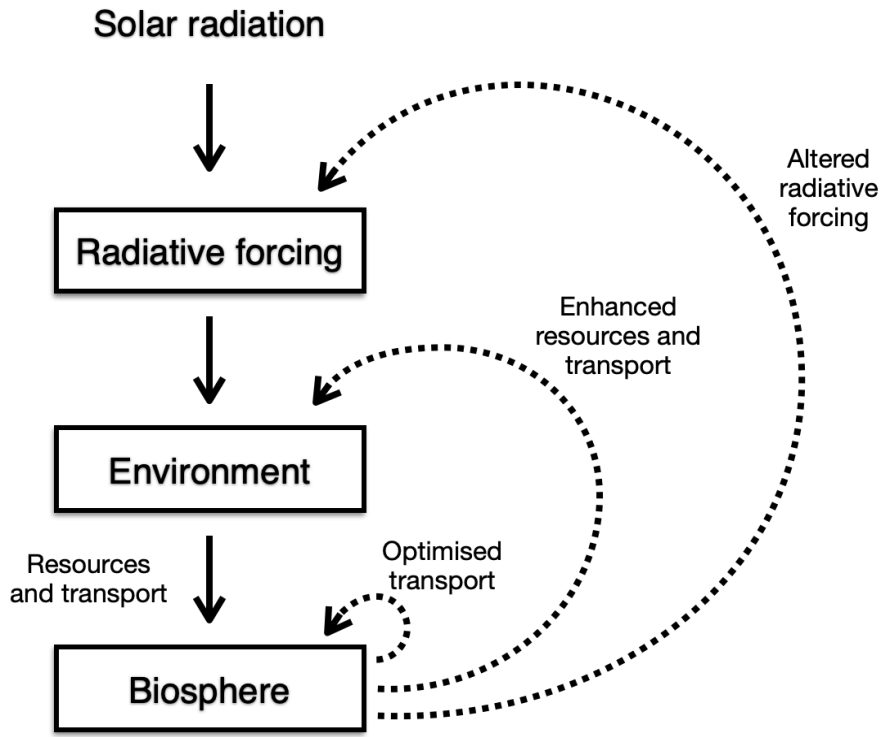

***Figure 5:*** *Life optimises transport limitations by different means: by optimising transport within their organisms (e.g., by fractal vascular networks and self-thinning laws), by enhancing resources and transport within the environment (e.g., by deeply reaching rooting systems and animals), and by altering the radiative forcing of the environment (e.g., by altering concentrations of carbon dioxide and oxygen).*

Applied to trees, these can be seen as a combination of two fractal networks connected by the trunk of the tree: (1) fractal root systems extract water and nutrients from the soil volume and bring these to the trunk; and (2) branching systems that connect the trunk with the leaves of the canopy that evaporate the water into the atmosphere while taking up carbon dioxide for photosynthesis. The fractal rooting systems maximise the contact area with the soil matrix, and the fractal canopies maximise the surface area for gas exchange with the atmosphere. As the mass exchange at the plant-environment interface takes place by diffusion, providing more contact area enhances the diffusive flux of the resources and thus enhances power and dissipative activity.

When trees are assembled in forests, the size distribution of individuals typically shows self-thinning scaling laws (e.g., Marquet et al. 2005). These can be seen as the outcome of minimising the metabolic, dissipative costs per unit mass and area for a given resource constraint (Enquist et al. 1999). As individual trees grow in size, their metabolism scales with the mass to the power of 3/4, as already mentioned above. A large tree thus dissipates more than a small tree, although this large tree is more efficient than a number of small trees with the same total mass. In other words, an old tree community with a few large individuals is more efficient than a young community with many little trees, even if they accomplish the same flux and dissipative activity. As this flux per unit area constrains the metabolic activity, what this then implies is that the associated mass increases, that is, it creates a greater disequilibrium with the same power with greater stand age.

These two examples reflect optimisation within the producers of the biosphere and do not involve enhancing resource availability and transport within the environment. One example of enhancing resources in the environment is given by the effect of the depth of the rooting zone on productivity (Kleidon and Heimann 1998, Kleidon 2023b). As plants trade water for carbon to be productive, water availability can constrain productivity during dry episodes and in arid regions. An imbalance between the supply of water by precipitation and evaporative demands is quite typical and impacts the ability to maintain continuous gas exchange. These imbalances can, however, be reduced or even compensated for by greater access to water stored in the soil, particularly in deeper soil layers. Plants can access this water by developing root systems that reach deep into the soil. In doing so, they are able to access water during dry episodes, maintain gas exchange, and remain productive. Ultimately, this utilisation of water stored in the soil is constrained by the climatological water balance of the region and the thermodynamic limit on evaporation. The effect of vegetation can nevertheless increase evaporation and productivity in the order of 10%, it yields evaporative fluxes consistent with observations (Kleidon and Heimann 1998, Kleidon 2023b), and it enhances hydrological cycling and precipitation on land (Kleidon and Heimann 1998). In other words, terrestrial vegetation appears to operate at the gas exchange limit set by thermodynamics, but with this limit being pushed to higher levels by enhancing water availability through deep-reaching rooting systems during dry episodes. By returning more water to the atmosphere, this effect enhances precipitation, thereby enhancing water availability on land.

Another mechanism for enhancing resources concerns the role of animals in the cycling of nutrients as resources. Apart from carbon and water, the biosphere needs other nutrients, such as nitrogen and phosphorus, to build biomass. These nutrients are typically in short supply, because they are energetically expensive to obtain, either by fixing nitrogen from air or by obtaining phosphorus from weathering rocks. Recycling these nutrients from existing biomass and making them available to the producers can be beneficial for producing new biomass, thus enhancing the power of the biosphere.

The presence of animals can result in an optimisation of nutrient availability for the producers that has been described by the "optimum grazing" hypothesis (McNaughton 1979). Herbivores can decompose biomass in their digestive systems better than soils by providing a continuously warm and humid environment that is conducive for decomposition. When nutrients are in short supply, these will eventually be taken up by the producers. Being locked up in biomass, the lack of nutrients in the environment can then limit the production of new biomass. When grazers take some of this biomass as their food (that is, as their energy source to maintain their metabolisms) and digest it, they release these nutrients back to the environment. These are then available again for the producers. Since the productivity of the ecosystem depends on the standing biomass of the producers (that is, more leaves, higher productivity) but also on nutrient availability (that is, more biomass, less nutrients available), there is an optimum level of grazing that can maximise the production of new biomass. As biomass is typically what feeds the consumers, a maximisation of biomass production would correspond to a maximisation of dissipation by the whole ecosystem of producers and consumers. The optimum grazing hypothesis can thus be seen as another manifestation of maximising the power of the biosphere.

Animals also actively enhance the transport of nutrients. Because they are able to physically move across space, they can perform work to move nutrients against physical gradients. Several studies have highlighted a significant role of animals, and food chains in general, in bringing nutrients back from the sea to the upstream producers (e.g., McClain and Naiman, 2008; Doughty et al., 2016; Buendia et al. 2018).  This is accomplished, e.g., by sea birds, who prey in the sea but rest and nest on land, or by migratory fish which breed in the river delta but migrate upstream, where they can serve as food for piscivores (fish-eating animals) that again live and breed on land.  The overall effect of these food chains is to move nutrients, particularly phosphorus, from downstream to upstream at quite significant rates.  This suggests that such food chains are not just dissipative systems, but play an important role in bringing back nutrients to the producers, thereby enhancing the power of the biosphere.

Last, but not least, life alters the radiative forcing of the planet, in terms of the absorption of solar radiation at the surface and in terms of the radiative properties of the atmosphere.  Vegetated surfaces typically reflect less sunlight, that is, they have a lower surface albedo.  A rainforest is a darker surface than bare ground.  This directly enhances the radiative heating by sunlight, enhancing buoyant transport and gas exchange.

Another example is given by the comparatively low concentrations of carbon dioxide in the atmosphere in Earth's recent past that are the reflection of the massive disequilibrium of reduced, organic carbon in biomass (and buried organic matter) and atmospheric oxygen that was generated by life.  The formation of this disequilibrium has substantially changed the composition of the atmosphere, and with this, its radiative properties.  As carbon dioxide is a relevant greenhouse gas, this change in composition has reduced the magnitude of the atmospheric greenhouse effect while this disequilibrium was generated throughout Earth's history.  It has thus transformed the radiative conditions of the planet under which work is performed and transport is generated, thus affecting the conditions for life to perform its work and its dissipative activities, and its ability to build up such massive disequilibrium.  As it requires work to remove carbon dioxide from the atmosphere - the work done by photosynthesis -, the low concentrations of carbon dioxide during the recent, interglacial past are likely also a manifestation of a biosphere operating at maximum power.

This list of examples is certainly not complete (see, e.g., the long history of optimality approaches in ecology, Lerdau et al., 2023).  I picked those examples to illustrate how existing concepts can be linked to a view of the biosphere as a dissipative system, and which aspects are likely to be optimised when the biosphere evolves towards maximising its power.  The means to do so vary across scales - from the optimised vascular networks within individuals at the local scale to the alteration of the radiative forcing at the planetary scale.  These means connect back to the dashed arrows shown in Figure 1, affecting the cycles and radiative forcing of the planet.

**5. Implications for the origins, habitability, and evolution of a life-dominated planet**

I have provided an update to the perspective described in Kleidon (2010) of life being part of the thermodynamic Earth system, being driven by power derived from sunlight and shaped by sequences of dissipative systems and their interactions with the forcing, as summarised by Figure 1.  Life as a dissipative system is embedded within these sequences and interactions.  It seems that it aims to evolve to maximise power, just like the heat engines of the climate system operate at maximum power.  This approach to the physical environment has been rather successful to describe and quantify the main features of the climate system and of global climate change.  This success suggests that a fuller application of this thermodynamic Earth system perspective to the biosphere can also result in such predictive outcomes.  A reflection of such predictability of the biosphere is reflected in well-established patterns by which productivity and typical vegetation types vary with climate.

From the examples I have described it seems that the key of applying this perspective to the biosphere lies in the role of material transport and exchange.  This is required to sustain metabolic reactions that are central to life, but it involves work needed to move the associated resources and metabolites around.  This aspect appears to be optimised by fractal networks that appear not just within living organisms, but also manifest themselves in, for instance, river systems (e.g., Rinaldo et al., 1992) and human-made infrastructure networks (e.g., Bettencourt, 2013).  It shifts the focus away from pure thermodynamics to indirect limitations by thermodynamics on generating the energy needed to move and to build such fractal structures.  It would seem that these aspects represent key aspects of developing this thermodynamic Earth system view further.

It would seem that this thermodynamic Earth system perspective could inform many questions about life and their planetary environments.  Not discussed here is the marine part of the biosphere.  Also there, mixing is known to represent a major constraint, setting the patterns of marine productivity.  This mixing is mostly wind-driven, that is, by the work done by the atmosphere.  This links this aspect to how much motion is generated, and thus to the hierarchy shown in Figure 1, but this aspect would require more work to specify.  One can also envision that this perspective informs theories of habitability, the origins of life, and the future evolution of the biosphere.  Habitability so far has mostly focused on the presence of liquid water (e.g., Seager 2013).  The thermodynamic Earth system perspective adds that it also needs to move, that is, the planet needs to perform work to maintain motion and the exchange of resources and metabolites.  This links directly to theories of the origins of life that focus on hydrothermal vents (Russell and Hall, 1997) - after all, these places are ones that do not just have geochemical energy available, but also moving water due to a geological heat source.  At a grander scale, this perspective links to the Gaia hypothesis of Lovelock and Margulis (1974), as homeostatic conditions in terms of carbon dioxide and temperature may result from maximising the power of the biosphere at geological time scales (Kleidon, 2023c).  When we also consider other planetary objects, one can use this perspective to distinguish planets and moons according to which types of work they can produce, and how much, so that they are able to sustain mass exchange and enable metabolic activities (Frank et al., 2017).

It would thus seem that the perspective laid out in Kleidon (2010) can provide profound further insights into the very basics of life as a planetary phenomenon that is deeply engrained in the overall functioning of the Earth system.  While the mechanisms seem specific to current Earth, the underlying principles of energy and mass conversions, limitations imposed by transport, possibilities for optimality and maximising power seem to be rather general so that these principles should be able to inform us about the future of the biosphere and life beyond Earth.